\documentclass[12pt]{article}
\usepackage{amsmath,epsf}

\renewcommand{\Re}{\operatorname{Re}}
\renewcommand{\Im}{\operatorname{Im}}
\newcommand{\Tr}{\operatorname{Tr}}
\newcommand{\dd}{\text{d}}

\begin{document}

\noindent ULM--TP/97-10 \\
October 1997
\vspace{3.0cm}

\centerline{\LARGE Billiard Systems in Three Dimensions:}
\vspace{0.3cm}
\centerline{\LARGE The Boundary Integral Equation and the Trace Formula}
\vspace{2.0cm}
\centerline{\large Martin Sieber}
\vspace{0.5cm}

\centerline{
Abteilung Theoretische Physik, Universit\"at Ulm, D-89069 Ulm, Germany}

\vspace{5.0cm}
\centerline{\bf Abstract}
\vspace{0.5cm}

We derive semiclassical contributions of
periodic orbits from a boundary integral
equation for three-dimensional billiard systems.
We use an iterative method that keeps track
of the composition of the stability matrix
and the Maslov index as an orbit is traversed. 
Results are given for isolated periodic orbits
and rotationally invariant families of periodic
orbits in axially symmetric billiard systems.
A practical method for determining the stability
matrix and the Maslov index is described.

\vspace{2.5cm}

\noindent PACS numbers: \\
\noindent 03.65.Sq ~ Semiclassical theories and applications. \\
\noindent 05.45.+b ~ Theory and models of chaotic systems.


\newpage

\section{Introduction}
\label{secintro}

Billiards are popular models for the study of dynamical 
systems and their quantum counterparts. They are
simpler than general systems with a potential and
exhibit a large spectrum of dynamical behaviour
ranging from integrability to chaoticity.
Until recently, most investigations have 
concentrated on two-dimensional billiards
whose numerical treatment requires much
less effort than higher-dimensional systems.
But two-dimensional systems also have rather
special properties like the division of
the energy surface into separate regions by 
invariant tori.

During the last three years there have been several
publications on three-dimensional billiard systems 
as models for chaotic systems 
\cite{PS95,Mat95,Hun95,AM96,AGHRRRSW96,Pro96a,Pro96b,Ste97,
BR97a,BR97b,HWG97,Pri97}. 
In other fields three-dimensional
billiard systems have served as models much longer,
like in nuclear or clusters physics.
An overview on these applications is given in \cite{BB97}.
In this context, Balian and Bloch investigated in detail 
semiclassical approximations for the level density in
three-dimensional billiard systems \cite{BB70,BB71,BB72}.
Their results for the contributions of periodic orbits
are expressed in terms of determinants of $2n$-dimensional
matrices, where $n$ is the number of reflection points of a
trajectory. In contrast to this, the results of Gutzwiller
for smooth three-dimensional systems are expressed in
terms of the $4 \times 4$-stability matrix \cite{Gut90},
and the relation between both results is not explicit.

For two dimensional billiard systems Harayama and Shudo
derived semiclassical contributions of isolated periodic
orbits in terms of the stability matrix starting from a
boundary integral equation \cite{HS92}. Their derivation involved
the transformation of $n$-dimensional matrices, where again
$n$ is the number of reflections of a trajectory.
For higher dimensional systems this method 
would be very elaborate. In addition, the relation
between the obtained index in the trace formula and
the number of conjugate points is not direct \cite{Bur95}.

We avoid this difficulty by applying an iterative
method that follows the trajectory from reflection
to reflection and keeps track of the composition
of the stability matrix and the Maslov index as
the trajectory is traversed. This method has 
proven to be convenient for the derivation of
uniform approximations for diffractive orbits
in two-dimensional billiard systems \cite{SPS97}.

The motivation for the present work is two-fold: 
firstly, we want to show how the contributions of isolated
periodic orbits in terms of the $4 \times 4$-stability
matrix can be obtained directly
from a boundary integral equation without
involving large matrices. Secondly,
we want to present a practical method for the
determination of semiclassical contributions of
isolated periodic orbits and families of orbits
in systems with axial symmetry, including the
determination of the stability matrix and the
Maslov index. Furthermore, the method
is convenient for the treatment of more 
complex contributions of periodic
orbits like those of bifurcations
that are of interest in nuclear physics applications
\cite{BRS97}.

\section{The boundary integral equation}
\label{secb}

A powerful method for investigating quantum billiards,
numerically as well as analytically, consists in
replacing the Schr\"odinger equation and boundary
conditions by an integral equation.
This integral equation is defined on the billiard boundary
and effectively reduces the dimension of the problem by one.
For three-dimensional billiards with Dirichlet boundary
conditions the integral equation has the following form \cite{KR74}
\begin{equation} \label{secb1}
- \int_{\partial {\cal D}} \! d \sigma \,
\partial_{\hat{n}\,'} \, G_0(\vec{r},\vec{r}\,',E) \, 
\partial_{\hat{n}} \psi(\vec{r}) =
\frac{1}{2} \partial_{\hat{n}\,'} \psi(\vec{r}\,') \; .
\end{equation}
Here ${\cal D}$ denotes the three-dimensional domain of the
billiard system, and $\partial {\cal D}$ is the boundary
surface. For simplicity we assume in the following that
the domain ${\cal D}$ is compact and the boundary
$\partial {\cal D}$ is ${\cal C}^2$.
The vectors $\vec{r}$ and $\vec{r}\,'$ are located
on the boundary $\partial {\cal D}$, and $d \sigma$ is
an area element at $\vec{r}$. Furthermore, $\hat{n}$
is a unit normal vector to the boundary pointing 
outwards, and $\partial_{\hat{n}} = \hat{n} \cdot \vec{\nabla}$
is the component of the gradient in direction of $\hat{n}$.
Dimensionless units with $\hbar = 2m = 1$ are used throughout,
$\psi$ is a solution of the Schr\"odinger equation with
Dirichlet boundary conditions, and $G_0$ denotes the free
outgoing Green function which is given by
\begin{equation} \label{secb2}
G_0(\vec{r},\vec{r}\,',E) =
- \frac{\exp \{ i k |\vec{r} - \vec{r}\,'| \}
}{4 \pi |\vec{r} - \vec{r}\,'|} \; ,
\end{equation}
where $k = \sqrt{E}$. Eq.\ (\ref{secb1}) is a Fredholm equation
of the second kind. It has non-trivial solutions only
if the determinant $\Delta(E) := \det(\boldsymbol{1} - \hat{Q}(E))$
vanishes, where $\hat{Q}$ is the integral operator defined 
on the boundary ${\partial \cal D}$ by
\begin{equation} \label{secb3}
\hat{Q} u(\vec{r}\,') = - 2 \int_{\partial {\cal D}} \! d \sigma \,
 \partial_{\hat{n}\,'} \, G_0(\vec{r},\vec{r}\,',E) \,
u(\vec{r}) \, .
\end{equation}

In recent years there were several publications on the Fredholm
determinant $\Delta(E)$ for two-dimensional billiards.
The integral equation in two dimensions is non-singular
if the boundary is continuously
differentiable, i.\,e.\ if it doesn't have corners or cusps
(see \cite{Rid79a,Rid79b}). For the following discussion
we assume that ${\cal D}$ is compact and simply connected,
and that the boundary $\partial {\cal D}$ is ${\cal C}^2$,
although the following results are valid under less
restrictive conditions. 
By an application of Fredholm theory, $\Delta(E)$ can be
represented by an absolutely convergent series which, when
approximated semiclassically, yields a series in
terms of so-called pseudo-orbits \cite{GP95a,Bur95,HST97}.
Considered as a function of $k$ the determinant $\Delta(E)$
is holomorphic in the $k$-plane except for a cut which
is due to the cut of the free outgoing Green function
in two dimensions.
It is known that in the upper half plane $\Im k \geq 0$ the
determinant has only zeros  which correspond to the eigenvalues
of the Schr\"odinger equation for the interior billiard with
Dirichlet boundary conditions $k = \pm \sqrt{E_n}$.
For $\Im k < 0$ there can be further zeros. These additional
zeros correspond to the resonances of the outside scattering
problem with Neumann boundary conditions \cite{BS95,THS97}.

The spectral staircase for the interior problem can be
represented in the following form 
\begin{equation} \label{secb4}
N(E) = N_{\text{sm}}(E) - \frac{1}{\pi}
\lim_{\varepsilon \rightarrow 0+} \Im \log
\frac{\Delta(E + i \varepsilon)}{\Delta(0)} \; ,
\end{equation}
where the second term on the right-hand side jumps by one at
every eigenvalue of the Schr\"odin\-ger equation (or by the
multiplicity of an eigenvalue if it is degenerate).
$N_{\text{sm}}(E)$ is a smooth function given by
\begin{equation} \label{secb5}
N_{\text{sm}}(E) = \frac{1}{\pi} \, \int_0^E \! dx \,
\lim_{\varepsilon \rightarrow 0+}
\Im \frac{\Delta'}{\Delta}(x+i\varepsilon) \; .
\end{equation}
It has been conjectured in \cite{BS95} and shown in \cite{EP97}
that $\pi N_{\text{sm}}(E)$ is equal to the total scattering
phase $\eta_{\text{N}}(E) = \frac{i}{2} \log \det S_{\text{N}}(E)$
for the outside scattering problem with Neumann boundary
conditions. The on-shell scattering matrix for this problem
is denoted by $S_{\text{N}}(E)$.

The asymptotic expansions of the smooth parts $\bar{N}(E)$
of the spectral staircase and $\bar{\eta}_{\text{N}}(E)/\pi$ 
of the total scattering phase agree in the first three leading
terms and in general differ in the next terms
\cite{BS95}. A relation between the complete asymptotic
expansions of the smooth part $\bar{N}(E)$
for the inner billiard problem and of the smooth
part $\bar{\eta}(E)$
for the outside scattering problem with the {\it same}
boundary conditions was conjectured and proved for
several leading terms (13 terms for the Dirichlet case)
in \cite{SU96}. It is stated that if the asymptotic
expansion of $\bar{N}(E)$ is
\begin{equation} \label{secb6}
\bar{N}(E) \sim a E + b E^{1/2} + c +
\sum_{n=1}^\infty c_n E^{1/2 - n} \, ,
\end{equation}
then the asymptotic expansion of $\bar{\eta}(E)$ is
\begin{equation} \label{secb7}
\frac{\bar{\eta}(E)}{\pi} 
\sim a E - b E^{1/2} + c - \sum_{n=1}^\infty c_n E^{1/2 - n} \; .
\end{equation}

A semiclassical evaluation of the second term on the
right-hand side of (\ref{secb4}) yields contributions
from interior periodic orbits as well as from periodic
orbits that run in the exterior of the billiard
(if they exist) \cite{Bur95,THS97}. The contributions from
the exterior orbits have to be cancelled by corresponding
contributions contained in $N_{\text{sm}}(E)$. This is
demonstrated in \cite{Hes97}.

Finally, the spectral determinant can be factorized in the form
$\Delta(E) = \Delta(0) \, D_{\text{int}}(E) \, D_{\text{ext}}(E)$
where $D_{\text{int}}(E)$ and $D_{\text{ext}}(E)$ are
completely specified in terms of the inside and outside
billiard problem, respectively \cite{BS95,THS97}.

In the present paper we do not aim at a 
generalization of the two-dimensional results to
three dimensions. We rather use the boundary
integral equation as starting
point for the derivation of semiclassical contributions
of periodic orbits. One difference in comparison to the
two-dimensional case is that the integral
equation in (\ref{secb1}) is singular since the
integral kernel has a $1/|\vec{r} - \vec{r}\,'|$
singularity. In the surface integral (\ref{secb1}) 
this singularity is integrable, but the trace of
the integral operator $\hat{Q}$ is infinite, and
the Fredholm determinant has to be regularized.

This can be done by considering the parameter dependent
spectral determinant
\begin{equation} \label{secb10}
\Delta_\lambda(E) := \det(\boldsymbol{1} - \lambda \hat{Q}(E))
= \exp \left\{ - \sum_{n=1}^\infty \frac{\lambda^n}{n}
\Tr \hat{Q}^n(E) \right\} \; , \qquad |\lambda| < R \; ,
\end{equation}
whose representation by the sum over $n$ has a positive
radius of convergence $R$ if all traces are finite.
If the first $l$ traces $\hat{Q}^n(E)$, $n=1,\dots,l$,
are infinite one can define a modified Fredholm determinant
by replacing the sum in (\ref{secb10}) by a similar expression
in which the first $l$ traces are missing \cite{FS83}. 
This expression can be analytically continued to
the whole complex $\lambda$-plane. The new determinant 
again has zeros at the eigenvalues of the Schr\"odinger
equation. In case of the integral operator (\ref{secb3})
the first two traces $\Tr \hat{Q}$ and $\Tr \hat{Q}^2$
are divergent.

\section{The semiclassical Green function}
\label{secc}

We take the derivative of eq.\,(\ref{secb4}) with
respect to $k$, insert (\ref{secb10}) for
$\lambda=1$, and obtain the following equation
for the level density in terms of the wavenumber $k$

\begin{equation} \label{secc1}
d(k) = d_{\text{sm},n_0}(k) + \frac{1}{\pi} \frac{d}{d k}  
\Im \sum_{n=n_0}^\infty \frac{1}{n} \Tr \hat{Q}^n (k)\; ,
\end{equation}
where
\begin{equation} \label{secc2}
\mbox{Tr}\, \hat{Q}^n (k) = (-2)^n \int_{\partial {\cal D}}
\! d \sigma_1 \dots d \sigma_n \;
\partial_{\hat{n}_1} G_0(\vec{r}_2,\vec{r}_1,E) \;
\partial_{\hat{n}_2} G_0(\vec{r}_3,\vec{r}_2,E) \dots
\partial_{\hat{n}_n} G_0(\vec{r}_1,\vec{r}_n,E) \; .
\end{equation}
Here the vectors $\vec{r}_i$ are located on the boundary,
$d \sigma_i$ is an area element at $\vec{r}_i$, and
$\hat{n}_i$ is the outward normal vector at $\vec{r}_i$.

As noted in the previous section, all traces of powers
of $\hat{Q}$ in the sum in (\ref{secc1}) are finite
if the sum starts with $n_0 \geq 3$. For the semiclassical
approximation this has the disadvantage that the
oscillatory contributions of $n$-bounce orbits with
$n < n_0$ are
shifted into the smooth part $d_{\text{sm},n_0}(k)$ through
eq.\ (\ref{secb5}) with the redefined spectral determinant.
This problem can be avoided by replacing the integral
kernel in (\ref{secb3}) by its leading semiclassical
term. Then the integral kernel is non-singular and
yields the same leading semiclassical contributions
to (\ref{secc2}) as before. For the derivation of
the semiclassical approximation, i.\,e.\ after
replacing the integral kernel by its leading
semiclassical term, $n_0$ thus can be set to $n_0=1$.

For real values of $k$
a second divergence occurs since the sum in
(\ref{secc1}) is in general divergent for real $k$.
One can avoid this by starting with an exact
formula for a smoothed level density \cite{BS95}.
We assume for the following derivation that $k$
has a sufficiently large imaginary part in order
to make the sum convergent, and that $\Re k > 0$.

The integrals that appear in eq.\ (\ref{secc2}) can be
given a simple interpretation in terms of the Green function
of the billiard system. According to the
multiple reflection expansion of Balian and Bloch \cite{BB70}
this Green function can be represented by
a sum over partial Green functions $G^{(n)}$ corresponding
to $n$ reflections on the billiard boundary, with $0 \leq n < \infty$
\begin{equation} \label{secc4}
G(\vec{r},\vec{r}\,',E) = \sum_{n=0}^\infty
G^{(n)}(\vec{r},\vec{r}\,',E) \; ,
\end{equation}
where
\begin{equation} \label{secc5}
G^{(n)}(\vec{r},\vec{r}\,',E) = (-2)^n \int_{\partial {\cal D}}
\! d \sigma_1 \dots d \sigma_n \;
G_0(\vec{r}_1,\vec{r}\,',E) \; \partial_{\hat{n}_1}
G_0(\vec{r}_2,\vec{r}_1,E) \dots \partial_{\hat{n}_n}
G_0(\vec{r},\vec{r}_n,E) \; .
\end{equation}
A comparison of (\ref{secc2}) and (\ref{secc5}) shows that
\begin{equation} \label{secc6}
d(k) = d_{\text{sm},n_0}(k) - \frac{2}{\pi} \, \frac{d}{d k} \Im
\sum_{n=n_0}^\infty \frac{1}{n} \int_{\partial {\cal D}} \! d \sigma \;
\partial_{\hat{n}'} G^{(n-1)}(\vec{r},\vec{r}\,',E)
|_{\vec{r}=\vec{r}\,'} \; .
\end{equation}

In the following we derive semiclassical contributions
of periodic orbits to $d(k)$ by evaluating the integrals
in (\ref{secc5}) and (\ref{secc6}) in stationary phase
approximation. The semiclassical evaluation of
(\ref{secc5}) expresses the partial Green function
$G^{(n)}$ in terms of classical trajectories
from $\vec{r}\,'$ to $\vec{r}$. These trajectories can
be of three kinds: interior orbits which run
inside the billiard domain ${\cal D}$, exterior orbits
which run outside ${\cal D}$, and ghost orbits
which cross the boundary and run inside and outside.
However, in the sum (\ref{secc4}) the leading order
contributions of ghost orbits from different $n$
cancel \cite{BB72}.
In the following we restrict to the consideration of
interior orbits, i.\,e.\ we denote by $G_{\text{sc}}^{(n)}$
the semiclassical contributions to $G^{(n)}$ from
interior orbits. We will show that this contribution
is given by
\begin{equation} \label{secc7}
G_{\text{sc}}^{(n)} (\vec{r},\vec{r}\,',E) = \sum_{\gamma_n}
\frac{1}{4 \pi \sqrt{|k^2 \det B_{\gamma_n}|}}
\exp \left\{ i k l_{\gamma_n}
- i \frac{\pi}{2} \nu_{\gamma_n} - i \pi \right\} \; ,
\end{equation}
where $\gamma_n$ labels all trajectories that run from
$\vec{r}\,'$ to $\vec{r}$ and are reflected $n$ times
on the boundary in between. Furthermore, $l_{\gamma_n}$
denotes the length of the trajectory $\gamma_n$, and
$\nu_{\gamma_n}$ is the number of conjugate points
from $\vec{r}\,'$ to $\vec{r}$ plus twice the number $n$
of reflections on the boundary. $B_{\gamma_n}$ is a
submatrix of the stability matrix for the trajectory.
The stability matrix is defined by
\begin{equation} \label{secc8}
\begin{pmatrix}        \dd \vec{q}_\perp \\
                       \dd \vec{p}_\perp \end{pmatrix} 
= M \, \begin{pmatrix} \dd \vec{q}_\perp\,' \\
                       \dd \vec{p}_\perp\,' \end{pmatrix}
\; , \qquad M = \begin{pmatrix} A & B \\ C & D \end{pmatrix} \; .
\end{equation}
Here $\dd \vec{q}_\perp\,'$ and $\dd \vec{p}_\perp\,'$ are
two-dimensional vectors which describe infinitesimal,
perpendicular deviations from a considered trajectory
at the starting point. $\dd \vec{q}_\perp$ and
$\dd \vec{p}_\perp$ are the corresponding
deviations at the end point that are obtained by a linearization
of the motion in the vicinity of the considered trajectory.
Correspondingly, $M$ is a $4 \times 4$-matrix, and $A$, $B$, $C$
and $D$ are $2 \times 2$-matrices. The determinant
of the submatrix $B$ is thus given by
$\det B = M_{13} M_{24} - M_{14} M_{23}$.
The denominator in (\ref{secc7}) is independent of $k$ since
$\det B$ is proportional to $1/E$.
Properties of the symplectic stability matrix are given
in appendix~\ref{apa}, and a method for its determination
is described in appendix~\ref{apb}.

We prove now eq.\ (\ref{secc7}) by induction.
For $n=0$ it is correct since 
\begin{equation} \label{secc9}
G_0(\vec{r},\vec{r}\,',E) =
G^{(0)}_{\text{sc}} (\vec{r},\vec{r}\,',E) \; .
\end{equation}
This follows from the form of $G_0$ in (\ref{secb2}) and the fact
that $\nu=0$ and $\det B = (\vec{r}-\vec{r}\,')^2$ for the free
motion (see the stability matrix for the free motion in 
appendix~\ref{apb}).

It remains to show that
\begin{equation} \label{secc10}
(-2) \int_{\partial {\cal D}} \! d \sigma \;
G_{\text{sc}}^{(n)}(\vec{r},\vec{r}_a,E) \;
\partial_{\hat{n}} G_0(\vec{r}_b,\vec{r},E)
\approx G^{(n+1)}_{\text{sc}} (\vec{r}_b,\vec{r}_a,E) \; ,
\end{equation}
where the approximate sign denotes that the integral
is evaluated in stationary phase approximation.
We use the following notation at the point $\vec{r}$
of the boundary: The momentum of the particle for the
incoming trajectory is denoted by $\vec{p}\,'$, and
the momentum of the outgoing trajectory is $\vec{p}$.
Both have modulus $k=\sqrt{E}$.

The normal derivative of the Green
function is given in leading semiclassical order by
\begin{equation} \label{secc10b}
\partial_{\hat{n}} G_0(\vec{r}_b,\vec{r},E) \approx - i\,
\hat{n} \cdot \vec{p} \; G_{\text{sc}}^{(0)}(\vec{r}_b,\vec{r},E) =
i \, k \, \cos \alpha \; G_{\text{sc}}^{(0)}(\vec{r}_b,\vec{r},E) \; ,
\end{equation}
where $\alpha$ is the angle between $-\hat{n}$ and $\vec{p}$.
The left-hand side of (\ref{secc10}) thus is a sum over
terms of the form
\begin{equation} \label{secc10c}
-2 i k \cos \alpha \int_{\partial {\cal D}} \!
d s_1 \, d s_2 \;
\frac{\exp \left\{ i k [l^{(n)} (\vec{r},\vec{r}_a)
+ l^{(0)} (\vec{r}_b,\vec{r}) ] 
- i \frac{\pi}{2} \nu^{(n)} \right\} }{
16 \pi^2 l^{(0)} \sqrt{|k^2 \det B^{(n)}|}} \; ,
\end{equation}
where the area element $d \sigma$ has been replaced by
$d s_1 \, d s_2$. 

The stationary points of the above integral are determined
by the conditions
\begin{equation} \label{secc11}
0 = \frac{d}{d s_{1,2}} [l^{(0)} (\vec{r}_b,\vec{r})
+ l^{(n)} (\vec{r},\vec{r}_a)]
= \hat{t}_{1,2} \cdot [ - \frac{\vec{p}}{k} +
\frac{\vec{p}\,'}{k}] \, ,
\end{equation}
where $t_1$ and $t_2$ are two orthogonal unit tangential
vectors at the considered point of reflection. 
In eq.\ (\ref{secc10c}) and in the following the length of a trajectory
is given two arguments when it is necessary to specify the
start and the end point of a trajectory. The second equation
in (\ref{secc11}) follows from the relation of the length
function to the action 
$S(\vec{r},\vec{r}\,',E) = k \, l(\vec{r},\vec{r}\,')$.

We neglect the possible solution $\vec{p} = \vec{p}\,'$ 
of (\ref{secc11}) since it leads to contributions
of ghost orbits. The other solution of (\ref{secc11})
yields the condition for elastic reflection:
$\vec{p}$, $\vec{p}\,'$ and $\hat{n}$ lie in one plane,
the reflection plane, and the angles between
$-\vec{p}$ and $\hat{n}$ and between $\hat{n}$ and
$\vec{p}\,'$ are identical. The value of this angle
$\alpha$ lies in the range $0 \leq \alpha < \pi/2$.
The sum over all stationary points thus expresses the
integral $I$ by a sum over all trajectories with $n+1$
reflections on the boundary.

For the following evaluations it will be useful to
express the tangential vectors and the normal vector
in local coordinate systems of the trajectory,
in which the first coordinate
is always in direction of the trajectory, and the
other two coordinates are perpendicular to it.
\begin{alignat}{6} \label{secc12}
\hat{n}    & = - && \cos \alpha \, \hat{e}_1   - \sin \alpha \, \hat{e}_2
          && =   && \cos \alpha \, \hat{e}_1'  - \sin \alpha \, \hat{e}_2'
          && =   && \cos \alpha \, \hat{e}_1'' - \sin \alpha \,
\left[ \cos \theta   \, \hat{e}_2'' + \sin \theta   \, \hat{e}_3''
\right] \, , \notag \\ 
\hat{t}_1  & =   && \sin \alpha \, \hat{e}_1   - \cos \alpha \, \hat{e}_2
          && =   && \sin \alpha \, \hat{e}_1'  + \cos \alpha \, \hat{e}_2'
          && =   && \sin \alpha \, \hat{e}_1'' + \cos \alpha \, 
\left[ \cos \theta   \, \hat{e}_2'' + \sin \theta   \, \hat{e}_3''
\right] \, , \notag \\
\hat{t}_2  & = && \hat{e}_3
          && = && \, \hat{e}_3'
          && = && \cos \theta \, \hat{e}_3'' - \sin \theta \, \hat{e}_2''
\; ,
\end{alignat}
where $\hat{t}_2 = \hat{n} \times \hat{t}_1$. In (\ref{secc12})
the unit vectors $\hat{e}_i''$ belong to the local coordinate
system for the incoming trajectory, and the unit vector $\hat{e}_2''$
lies in the reflection plane of the previous reflection. This
coordinate system is rotated around the trajectory such
that the new vector $\hat{e}_2'$ lies in the new reflection
plane. The unprimed coordinate system is the local coordinate
system directly after the reflection, and $\hat{e}_2$ again
lies in the reflection plane.

We continue now with the evaluation of the integral in
eq.\ (\ref{secc10c}) by stationary phase approximation.
For that purpose the second derivatives of the exponent
have to be determined. According to (\ref{secc11}) they
consist of two parts, the derivatives of the tangential
vectors and the derivatives of the momenta. Those
of the tangential vectors follow from eq.\,(\ref{apd6})
of appendix \ref{apd} and are given by
\begin{equation} \label{secc13}
\frac{\partial \hat{t}_1}{\partial s_1} = - \frac{\hat{n}}{R_a}
\, , \qquad
\frac{\partial \hat{t}_1}{\partial s_2} = - \frac{\hat{n}}{R_c}
\, , \qquad
\frac{\partial \hat{t}_2}{\partial s_1} = - \frac{\hat{n}}{R_c}
\, , \qquad
\frac{\partial \hat{t}_2}{\partial s_2} = - \frac{\hat{n}}{R_b} \, ,
\end{equation}
where $R_a$, $R_b$ are the radii of curvature of the
surface in the reflection plane and perpendicular
to it, respectively. $R_a$, $R_b$ and $R_c$ are given in
(\ref{apd4}) in terms of the two main radii of curvature
$R_1$ and $R_2$ and the angle $\beta$ between $\hat{t}_1$
and the main curvature direction for $R_1$.

The derivatives of the momenta of the incoming and outgoing
trajectory are expressed in terms of the local coordinate
systems before and after the reflection, respectively.
\begin{alignat}{5} \label{secc15}
\frac{\partial \vec{p}    }{\partial s_1} & =
(\hat{t}_1 \cdot \vec{\nabla}) \vec{p} && =
- && \cos \alpha \,  \left(
\frac{\partial p_2}{\partial q_2} \hat{e}_2
+ \frac{\partial p_3}{\partial q_2} \hat{e}_3 \right)
\, , & \qquad
\frac{\partial \vec{p}    }{\partial s_2} & =
(\hat{t}_2 \cdot \vec{\nabla}) \vec{p} && = 
\frac{\partial p_2}{\partial q_3} \hat{e}_2
+ \frac{\partial p_3}{\partial q_3} \hat{e}_3 \, ,
\notag \\
\frac{\partial \vec{p}\,' }{\partial s_1} & =
(\hat{t}_1 \cdot \vec{\nabla}) \vec{p}\,' && =
  && \cos \alpha \,  \left( 
\frac{\partial p_2'}{\partial q_2'}
\hat{e}_2'
+ \frac{\partial p_3'}{\partial q_2'} \hat{e}_3' \right)
\, , & \qquad
\frac{\partial \vec{p}\,' }{\partial s_2} & =
(\hat{t}_2 \cdot \vec{\nabla}) \vec{p}\,' && = 
\frac{\partial p_2'}{\partial q_3'} \hat{e}_2'
+ \frac{\partial p_3'}{\partial q_3'} \hat{e}_3' \, ,
\end{alignat}
and with (\ref{secc11}), (\ref{secc13}) and (\ref{secc15})
the second derivatives of the length function follow as
\begin{alignat}{2} \label{secc16}
a_1 & := \frac{\partial^2}{\partial s_1 \partial s_1} \left[ 
l^{(0)}(\vec{r}_b,\vec{r}) + l^{(n)}(\vec{r},\vec{r}_a) 
\right] &&=
- \frac{2 \cos \alpha}{R_a} + \left[
- \frac{\partial p_2 }{\partial q_2 }
+ \frac{\partial p_2'}{\partial q_2'} \right]
\, \frac{\cos^2 \alpha}{k} \, ,
\nonumber \\
b_1 & := \frac{\partial^2}{\partial s_2 \partial s_1} \left[
l^{(0)}(\vec{r}_b,\vec{r}) + l^{(n)}(\vec{r},\vec{r}_a) 
\right] &&=
- \frac{2 \cos \alpha}{R_c} + \left[
  \frac{\partial p_2 }{\partial q_3 }
+ \frac{\partial p_2'}{\partial q_3'} \right]
\, \frac{\cos \alpha}{k} \, ,
\nonumber \\
c_1 & := \frac{\partial^2}{\partial s_1 \partial s_2} \left[
l^{(0)}(\vec{r}_b,\vec{r}) + l^{(n)}(\vec{r},\vec{r}_a)
\right] &&=
- \frac{2 \cos \alpha}{R_c} + \left[
  \frac{\partial p_3 }{\partial q_2 }
+ \frac{\partial p_3'}{\partial q_2'} \right]
\, \frac{\cos \alpha}{k} \, ,
\nonumber \\
d_1 & := \frac{\partial^2}{\partial s_2 \partial s_2} \left[
l^{(0)}(\vec{r}_b,\vec{r}) + l^{(n)}(\vec{r},\vec{r}_a)
\right] &&=
- \frac{2 \cos \alpha}{R_b} + \left[
- \frac{\partial p_3 }{\partial q_3 }
+ \frac{\partial p_3'}{\partial q_3'} \right]
\, \frac{1}{k} \, .
\end{alignat}
We discuss in more detail how the partial derivatives
of the momenta in (\ref{secc16}) have to be interpreted.
The derivatives
$\partial p_i / \partial q_j$ describe the change of
the component $p_i$ upon a change of $q_j$ when the
other components of $\vec{q}$ and $\vec{q}_b$ are
hold fixed. Thus it can be recognized as an element
of the $N$-matrix of appendix \ref{apa} for the
free motion from $\vec{q}$ to $\vec{q}_b$. By the
relations of appendix \ref{apa} it can be expressed
in terms of the stability matrix $M_T$ for this part
of the trajectory. Note, however, that primed and
unprimed quantities have a different meaning here
and in appendix \ref{apa}. Here the unprimed quantities
are the quantities directly after the reflection,
i.\,e.\ they describe the beginning of the trajectory
from $\vec{q}$ to $\vec{q}_b$. In the appendix \ref{apa},
however, the unprimed quantities are those at the
end of a trajectory, so the meaning is reversed.

The derivatives $\partial p_i' / \partial q_j'$ 
on the other hand describe the change of
the component $p_i'$ upon a change of $q_j'$ when the
other components of $\vec{q}\,'$ and $\vec{q}_a$ are
hold fixed. Again these partial derivatives are 
elements of an $N$-matrix, but now for the trajectory
from $\vec{q}_a$ to $\vec{q}\,'$, including the final
rotation that transforms the double primed coordinate
system into the primed one. Again by the relations of
appendix \ref{apa} they can be expressed in terms of
the stability matrix $M'= M_S \, M^{(n)}$, where
the rotation matrix $M_S$ is given in (\ref{apb4}). 

The quantities $a_1, \dots ,d_1$ in (\ref{secc16}) can
also be expressed in terms of the stability matrix for the
full trajectory from $\vec{q}_a$ to $\vec{q}_b$, that
is given by $M := M^{(n+1)} = M_T \, M_R \, M_S \, M^{(n)}$
where $M_R$ is the stability matrix for a reflection
given in (\ref{apb5}).
The following expressions can be proved by a direct although
lengthy evaluation that was done with Maple. One finds that
\begin{alignat}{4} \label{secc17}
a_1 & = - && \frac{[M_{13}  \, M_{24}' - M_{14}  \, M_{23}']
              }{[M_{13}' \, M_{24}' - M_{14}' \, M_{23}']}
\,      \frac{\cos^2 \alpha}{l^{(0)}}
\, , \qquad &
b_1 & =   && \frac{[M_{13}  \, M_{14}' - M_{14}  \, M_{13}']
              }{[M_{13}' \, M_{24}' - M_{14}' \, M_{23}']}
\,      \frac{\cos \alpha}{l^{(0)}}
\, , \notag \\
c_1 & =   && \frac{[M_{23}  \, M_{24}' - M_{24}  \, M_{23}']
              }{[M_{13}' \, M_{24}' - M_{14}' \, M_{23}']}
\,      \frac{\cos \alpha}{l^{(0)}}
\, , \qquad &
d_1 & = - && \frac{[M_{23}  \, M_{14}' - M_{24}  \, M_{13}']
              }{[M_{13}' \, M_{24}' - M_{14}' \, M_{23}']}
\,      \frac{1}{l^{(0)}} \, .
\end{alignat}
and 
\begin{equation} \label{secc18}
a_1 \, d_1 - b_1 \, c_1 = 
-        \frac{[M_{13}  \, M_{24}  - M_{14}  \, M_{23} ]
             }{[M_{13}' \, M_{24}' - M_{14}' \, M_{23}']}
\, \frac{\cos^2 \alpha}{(l^{(0)})^2}
= - \frac{\cos^2 \alpha \, \det B}{(l^{(0)})^2 \, \det B'} \, ,
\end{equation}
where $M = M_T \, M_R \, M'$ and $M' = M_S \, M^{(n)}$ as before.
With (\ref{secc18}) the contribution of a stationary point to
the integral in (\ref{secc10c}) can be evaluated and results in
\begin{align} \label{secc19}
I_{\text{sp}} & = -i k \cos \alpha \int \! d s_1 \, d s_2 \;
\frac{\exp \left\{ i k [l^{(n)}+l^{(0)}] 
+ \frac{ik}{2} [a_1 s_1^2 + (b_1 + c_1) s_1 s_2 + d_1 s_2^2]
- i \frac{\pi}{2} \nu^{(n)}
\right\} }{8 \pi^2 l^{(0)} \sqrt{|k^2 \det B'|}} 
\notag \\
& = \frac{\exp \left\{ i k l^{(n+1)} 
- i \frac{\pi}{2} \nu^{(n+1)} - i \pi \right\}
}{4 \pi \sqrt{|k^2 \det B^{(n+1)}|}} \, ,
\end{align}
where $\det B' = \det B^{(n)}$ has been used, and
\begin{equation} \label{secc20}
\nu^{(n+1)} = \nu^{(n)} + 2 +
\begin{cases}
0 & \text{if $a_1 > 0$ and $(a_1 \, d_1 - b_1 \, c_1) > 0$} \, , \\
1 & \text{if $(a_1 \, d_1 - b_1 \, c_1) < 0$}               \, , \\
2 & \text{if $a_1 < 0$ and $(a_1 \, d_1 - b_1 \, c_1) > 0$} \, .
\end{cases}
\end{equation}

The expression (\ref{secc19}) has the correct form of a
contribution to the semiclassical Green function in (\ref{secc7}).
It remains to show that the condition (\ref{secc20}) gives
the correct index $\nu^{(n+1)}$ that is
defined as the number of conjugate points from $\vec{r}_a$ 
to $\vec{r}_b$ plus twice the number of reflections in between.

The index $\nu^{(n+1)}$ increases in comparison to $\nu^{(n)}$
by 2 because of the additional reflection, and possibly further
by 0, 1 or 2, depending on the number of conjugate points between
$\vec{r}$ and $\vec{r}_b$. To determine these conjugate points
we consider $\det B$ as a function of the length $l^{(0)}$
as we vary the end point of the trajectory from $\vec{r}$ to
$\vec{r}_b$. A conjugate point occurs at every zero
of $\det B$, and there can be at most two zeros between $\vec{r}$ to
$\vec{r}_b$, because $\det B$ is a quadratic function of $l^{(0)}$.
This follows from the form of the matrix for free motion $M_T$ in
(\ref{apb4}). We note that a multiplication of a stability
matrix by a reflection matrix $M_R$ changes only the sign of $\det B$.

From these considerations follows that there is exactly one zero
of $\det B$ between $\vec{r}$ and $\vec{r}_b$ if
$(a_1 \, d_1 - b_1 \, c_1) < 0$ because then $\det B$ has 
changed sign between $\vec{r}$ and $\vec{r}_b$. If
$(a_1 \, d_1 - b_1 \, c_1) > 0$ there can be no or two zeros
of $\det B$ between $\vec{r}$ and $\vec{r}_b$. This 
depends on the sign of $a_1$.
The quantity $a_1$ has at most one zero as a function of
$l^{(0)}$ as can be seen from its dependence on $l^{(0)}$ in 
\begin{equation} \label{secc20b}
a_1 = \frac{\cos^2 \alpha}{l^{(0)}} + e_1 
\, , \qquad
b_1 = e_2 
\, , \qquad
c_1 = e_2 
\, , \qquad
d_1 = \frac{1}{l^{(0)}} + e_3 \, ,
\end{equation}
where $e_1$, $e_2$ and $e_3$ are independent of $l^{(0)}$.
Furthermore, the dependence of the coefficient $a_1, \dots ,d_1$
on $l^{(0)}$ shows that a zero of $a_1$ is
always after a first zero of $\det B$, and before a second
zero of $\det B$ there is always a zero of $a_1$. (If $a_1=0$
then $(a_1 \, d_1 - b_1 \, c_1) < 0$ and for that reason
there must be a zero of $(a_1 \, d_1 - b_1 \, c_1)$ before
since both, $a_1$ and $(a_1 \, d_1 - b_1 \, c_1)$, are positive
for small $l^{(0)}$. Furthermore, $a_1$ and $d_1$
are monotonously decreasing functions of  $l^{(0)}$. If 
$a_1 \, d_1 = e_2^2 $ for two different values of $l^{(0)}$, both,
$a_1$ and $d_1$, must have changed sign in between.)
Thus the number of conjugate points is given
correctly by (\ref{secc20}) and the semiclassical expression
for $G^{(n)}$ in (\ref{secc7}) is correct. 

\section{Contributions of periodic orbits}
\label{secd}

We continue with the evaluation of the integral (\ref{secc6})
for the spectral density. Inserting the semiclassical
expressions (\ref{secc10b}) and (\ref{secc7}) for the
partial Green function into
(\ref{secc6}) yields contributions to the level
density of the form
\begin{equation} \label{secd1}
\frac{2}{\pi} \, \frac{d}{d k} \Im \frac{1}{n}
\int_{\partial {\cal D}} \! d s_1 \, d s_2 \;
i \, k \cos \alpha \,
\frac{\exp \left\{ i k l^{(n-1)} - i \frac{\pi}{2} \nu^{(n-1)}
\right\} }{4 \pi \sqrt{|k^2 \det B^{(n-1)}|}} \, .
\end{equation}

We consider now the contributions of stationary
points to the integral in (\ref{secd1}). The
stationary points are determined by
\begin{equation} \label{secd2}
0 = \frac{d}{d s_{1,2}} [l^{(n-1)} (\vec{r},\vec{r})]
= \hat{t}_{1,2} \cdot [ - \frac{\vec{p}}{k} +
\frac{\vec{p}\,'}{k}] \, ,
\end{equation}
which is the condition for a specular reflection.
Again $\vec{p}\,'$ is the momentum directly before
the reflection, and $\vec{p}$ the momentum directly
after the reflection.
The contributions thus come from periodic orbits
with $n$ specular reflections on the billiard wall.
We consider one such orbit and label it by $\gamma$.
It has $n/r_\gamma$ distinct reflections on the
boundary if $r_\gamma$ is its repetition number.
Consequently, one obtains 
$n/r_\gamma$ identical semiclassical contributions for
this orbit, since $n/r_\gamma$ different stationary
points correspond to it. (That the contributions
are identical follows from the fact that a stationary
phase contribution to $\Tr \hat{Q}^n$ in (\ref{secc2})
does not depend on the order in which the $n$ integrals
are evaluated.)

For a determination of the second derivatives of the
exponent in (\ref{secd1}) the derivatives of the
momenta $\vec{p}$ and $\vec{p}\,'$ are needed.
\begin{alignat}{2} \label{secd3}
\frac{\partial \vec{p}    }{\partial s_1} & =
(\hat{t}_1 \cdot \vec{\nabla}) \vec{p} && =
\left(
- \frac{\partial p_2}{\partial q_2 } \hat{e}_2
- \frac{\partial p_3}{\partial q_2 } \hat{e}_3
+ \frac{\partial p_2}{\partial q_2'} \hat{e}_2
+ \frac{\partial p_3}{\partial q_2'} \hat{e}_3 \right) \cos \alpha \, ,
\notag \\
\frac{\partial \vec{p}    }{\partial s_2} & =
(\hat{t}_2 \cdot \vec{\nabla}) \vec{p} && = 
  \frac{\partial p_2}{\partial q_3 } \hat{e}_2
+ \frac{\partial p_3}{\partial q_3 } \hat{e}_3
+ \frac{\partial p_2}{\partial q_3'} \hat{e}_2
+ \frac{\partial p_3}{\partial q_3'} \hat{e}_3 \, ,
\notag \\
\frac{\partial \vec{p}\,' }{\partial s_1} & =
(\hat{t}_1 \cdot \vec{\nabla}) \vec{p}\,' && =
  \left( 
- \frac{\partial p_2'}{\partial q_2 } \hat{e}_2'
- \frac{\partial p_3'}{\partial q_2 } \hat{e}_3'
+ \frac{\partial p_2'}{\partial q_2'} \hat{e}_2'
+ \frac{\partial p_3'}{\partial q_2'} \hat{e}_3' \right) \cos \alpha \, ,
\notag \\
\frac{\partial \vec{p}\,' }{\partial s_2} & =
(\hat{t}_2 \cdot \vec{\nabla}) \vec{p}\,' && = 
  \frac{\partial p_2'}{\partial q_3 } \hat{e}_2'
+ \frac{\partial p_3'}{\partial q_3 } \hat{e}_3'
+ \frac{\partial p_2'}{\partial q_3'} \hat{e}_2'
+ \frac{\partial p_3'}{\partial q_3'} \hat{e}_3' \, .
\end{alignat}
In contrast to the results in (\ref{secc15}) there are
additional terms in (\ref{secd3}). They arise because
a change of the reflection point changes the initial
point as well as the end point of the trajectory from
$\vec{r}$ to $\vec{r}$. With (\ref{secd2}), (\ref{secc13})
and (\ref{secd3}) we obtain the following expressions
\begin{alignat}{3} \label{secd4}
a_2 & := & \frac{\partial^2}{\partial s^2_1} &
\left[ l^{(n-1)}(\vec{r},\vec{r}) \right] &&=
- \frac{2 \cos \alpha}{R_a} + \left[
- \frac{\partial p_2 }{\partial q_2 }
+ \frac{\partial p_2 }{\partial q_2'}
- \frac{\partial p_2'}{\partial q_2 }
+ \frac{\partial p_2'}{\partial q_2'} \right] 
\frac{\cos^2 \alpha}{k} \, ,
\nonumber \\
b_2 & := & \frac{\partial^2}{\partial s_2 \partial s_1} &
\left[ l^{(n-1)}(\vec{r},\vec{r}) \right] &&=
- \frac{2 \cos \alpha}{R_c} + \left[
  \frac{\partial p_2 }{\partial q_3 }
+ \frac{\partial p_2 }{\partial q_3'}
+ \frac{\partial p_2'}{\partial q_3 }
+ \frac{\partial p_2'}{\partial q_3'} \right]
\frac{\cos \alpha}{k} \, ,
\nonumber \\
c_2 & := & \frac{\partial^2}{\partial s_1 \partial s_2} &
\left[ l^{(n-1)}(\vec{r},\vec{r}) \right] &&=
- \frac{2 \cos \alpha}{R_c} + \left[
  \frac{\partial p_3 }{\partial q_2 }
- \frac{\partial p_3 }{\partial q_2'}
- \frac{\partial p_3'}{\partial q_2 }
+ \frac{\partial p_3'}{\partial q_2'} \right]
\frac{\cos \alpha}{k} \, ,
\nonumber \\
d_2 & := & \frac{\partial^2}{\partial s^2_2} &
\left[ l^{(n-1)}(\vec{r},\vec{r}) \right] &&=
- \frac{2 \cos \alpha}{R_b} + \left[
- \frac{\partial p_3 }{\partial q_3 }
- \frac{\partial p_3 }{\partial q_3'}
+ \frac{\partial p_3'}{\partial q_3 }
+ \frac{\partial p_3'}{\partial q_3'} \right] \frac{1}{k} \, .
\end{alignat}
Again all partial derivatives of the momenta are elements
of an $N$-matrix that corresponds to the stability matrix
$M' = M_S \, M^{(n-1)}$, and thus the quantities
$a_2, \dots ,d_2$ can be expressed in terms of the elements
of $M'$. They can further be expressed in terms of the
full stability matrix of the orbit $M = M_\gamma = M_R \, M'$,
where $M_R$ describes the reflection. One finds that
\begin{alignat}{4} \label{secd5}
a_2 & =   && \frac{ {\cal M}_{42}}{k \det B} \cos^2 \alpha
\; , \qquad \qquad & 
b_2 & = - && \frac{ {\cal M}_{41}}{k \det B} \cos \alpha
\; , \notag \\
c_2 & = - && \frac{ {\cal M}_{32}}{k \det B} \cos \alpha
\; , \qquad \qquad & 
d_2 & =   && \frac{ {\cal M}_{31}}{k \det B} \, ,
\end{alignat}
and 
\begin{equation} \label{secd6}
a_2 \, d_2 - b_2 \, c_2 = \frac{\det (M-1) 
\cos^2 \alpha}{k^2 \det B} \; .
\end{equation}
The quantity ${\cal M}_{ij}$ in (\ref{secd5}) is the determinant
of the $3 \times 3$-matrix that is obtained by dropping the
$i$-th row and the $j$-th column of the matrix $(M-1)$.

With these expressions a stationary phase approximation
for the contribution of an isolated periodic orbit results in
\begin{align} \label{secd7}
d_\gamma(k) & = - \frac{l^{(n-1)} \cos \alpha}{2 \pi^2 r_\gamma}
\, \Im \int \! \! d s_1 \, d s_2 \;
\frac{k \exp \left\{ i k l^{(n-1)} 
+ \frac{ik}{2} (a_2 s_1^2 + (b_2 + c_2) s_1 s_2 + d_2 s_2^2)
- i \frac{\pi}{2} \nu^{(n-1)} \right\} }{\sqrt{|k^2 \det B^{(n-1)}|}}
\notag \\
& =  \frac{l_\gamma}{\pi r_\gamma \, \sqrt{|\det (M_\gamma - 1)|}}
\Re \exp \left\{ i k l_\gamma - \frac{i \pi}{2} \mu_\gamma \right\} \, ,
\end{align}
where the relations $|\det B^{(n-1)}| = |\det B|$,
$l_\gamma = l^{(n-1)}$ and
\begin{equation} \label{secd8}
\mu_\gamma = \nu^{(n-1)} + 2 +
\begin{cases}
0 & \text{if $a_2 > 0$ and $(a_2 \, d_2 - b_2 \, c_2) > 0$} \, , \\
1 & \text{if $(a_2 \, d_2 - b_2 \, c_2) < 0$}               \, , \\
2 & \text{if $a_2 < 0$ and $(a_2 \, d_2 - b_2 \, c_2) > 0$} \, ,
\end{cases}
\end{equation}
have been used.
Eq.\ (\ref{secd7}) is the expected contribution of an isolated
periodic orbit. The relations for determining the Maslov index
$\mu$ of a periodic orbit are summarized in appendix~\ref{apc}.
The conditions for the additional contributions to $\mu_\gamma$
in (\ref{secd8}) can be shown to be identical to the general
conditions in \cite{CRL90}.

\bigskip

In applications three-dimensional
billiard systems often have an axial symmetry. Then periodic
orbits appear in one-parameter families. In the following we
derive the contribution of such a family. A general treatment of
semiclassical trace formulas in the presence of continuous
symmetries can be found in \cite{CL91}. In this article,
a general formula is given for families of orbits in axially
symmetric systems.
The formula which is obtained in the following has a different
form which is not explicitly independent of the starting
point of the trajectory, but it has the advantage that it
is expressed directly in terms of the stability matrix $M$.
Of course, both formulas must yield the same result.

In presence of an axial symmetry the quantity $(a_2 \, d_2 - b_2 \, c_2)$
vanishes and the symmetric matrix $\left( \begin{smallmatrix} a_2 & b_2 \\
c_2 & d_2 \end{smallmatrix} \right)$ has eigenvalues 0 and 
$(a_2 + d_2)$. By a rotation of the coordinate system the
matrix can be diagonalized and one obtains
\begin{align} \label{secd9}
d_\gamma(k) & = - \frac{l^{(n-1)} \cos \alpha}{2 \pi^2 r_\gamma}
\, \Im \int \! d s_1' \, d s_2' \;
\frac{k \exp \left\{ i k l^{(n-1)} + \frac{ik}{2} (a_2+d_2) {s_1'}^2
- i \frac{\pi}{2} \nu^{(n-1)} \right\} }{\sqrt{|k^2 \det B^{(n-1)}|}}
\notag \\
& =  \frac{l_\gamma \rho \cos \alpha}{\pi r_\gamma \, N} \Re \left[
\sqrt{\frac{2 \pi k}{|k {\cal M}_{42} \cos^2 \alpha + k {\cal M}_{31}|}}
\exp \left\{ i k l_\gamma - \frac{i \pi}{2} \mu_\gamma 
- \frac{i \pi}{4} \right\} \right] \, ,
\end{align}
where $l_\gamma = l^{(n-1)}$ and
\begin{equation} \label{secd10}
\mu_\gamma = \nu^{(n-1)} + 2 +
\begin{cases}
0 & \text{if $(a_2+d_2) > 0$} \, , \\
1 & \text{if $(a_2+d_2) < 0$} \, .
\end{cases}
\end{equation}

In (\ref{secd9}) the integral over $s_2'$ was evaluated over
the range $2 \pi \rho/N$ where $\rho$ is the distance of the
stationary point on the billiard surface to the symmetry axis,
and $N$ is the number of distinct rotations 
around the symmetry axis that map the orbit onto itself.
The result (\ref{secd9}) does not depend on the reflection
point that is chosen as starting point of the trajectory,
although this is not explicit from its form.
The reason is again, that a semiclassical contribution to
$\Tr \hat{Q}^n$ in (\ref{secc2}) does not depend on the
order in which the $n$ integrals are evaluated.
\bigskip

We acknowledge financial support by the Deutsche
Forschungsgemeinschaft under contract No.\ DFG-Ste
241/6-1 and /7-2.

\appendix

\section{General properties of the stability matrix}
\label{apa}

The stability matrix $M$ is defined by the following relation
\begin{equation} \label{apa1}
\begin{pmatrix}
\dd q_2 \\ \dd q_3 \\ \dd p_2 \\ \dd p_3
\end{pmatrix} = M \, \begin{pmatrix}
\dd q_2' \\ \dd q_3' \\ \dd p_2' \\ \dd p_3'
\end{pmatrix}
\; , 
\qquad M = \begin{pmatrix}
M_{11} & M_{12} & M_{13} & M_{14} \\
M_{21} & M_{22} & M_{23} & M_{24} \\
M_{31} & M_{32} & M_{33} & M_{34} \\
M_{41} & M_{42} & M_{43} & M_{44}
\end{pmatrix} \; .
\end{equation}
Here $\dd q_2'$, $\dd q_3'$, $\dd p_2'$, and $\dd p_3'$ are
infinitesimal deviations from the considered trajectory at
the starting point. They are given in a local coordinate
system in which the first coordinate is along the trajectory,
and the other two coordinates are perpendicular to it. 
The unprimed quantities $\dd q_2$, $\dd q_3$, $\dd p_2$
and $\dd p_3$ are the corresponding infinitesimal deviations
at the end point of the trajectory. They follow from a
linearization of the motion in the vicinity of the considered
trajectory. In a compact notation
\begin{equation} \label{apa2}
\begin{pmatrix}
\dd \vec{q}_\perp \\ \dd \vec{p}_\perp
\end{pmatrix} 
= M \, \begin{pmatrix}
\dd \vec{q}_\perp\,' \\ \dd \vec{p}_\perp\,'
\end{pmatrix}
\, , \qquad
M = \begin{pmatrix} A & B \\ C & D \end{pmatrix} \, ,
\end{equation}
where $A$, $B$, $C$ and $D$ are $2 \times 2$-matrices.

The stability matrix is a symplectic matrix. It satisfies
$M^T \, J \, M = J$, where $J$ denotes the matrix
$J = \left( \begin{smallmatrix} 0 & I \\ -I & 0
\end{smallmatrix} \right)$ and $I$ is the two-dimensional
unity matrix. The inverse of $M$ is given by
$M^{-1} = - J \, M^T \, J$, and from 
$M^{-1} \, M = M \, M^{-1} = 1$ one obtains the following
relations between the submatrices of $M$
\begin{equation} \label{apa3}
\begin{array}{ccc}
A^T \, C = C^T \, A
\; , \quad &
B^T \, D = D^T \, B
\; , \quad &
A^T \, D - C^T \, B = I \, , \\
A \, B^T = B \, A^T
\; , \quad &
C \, D^T = D \, C^T
\; , \quad &
A \, D^T - B \, C^T = I \, .
\end{array}
\end{equation}
Both rows of equations are equivalent to each other and
to the symplectic condition for $M$.
Altogether they yield 6 independent conditions for the 
matrix elements of $M$, so that only 10 elements of $M$
are independent. 

Another useful quantity is the matrix $N$ which is defined in
the following. It is obtained by solving the linear equations
in (\ref{apa2}) for $\vec{p}_\perp$ and $\vec{p}_\perp\,'$
\begin{equation} \label{apa4}
       \begin{pmatrix} \dd \vec{p}_\perp\,' \\
                       \dd \vec{p}_\perp \end{pmatrix}
= N \, \begin{pmatrix} \dd \vec{q}_\perp\,' \\
                       \dd \vec{q}_\perp \end{pmatrix}
=      \begin{pmatrix} \tilde{A} & \tilde{B} \\
                       \tilde{C} & \tilde{D} \end{pmatrix} \,
       \begin{pmatrix} \dd \vec{q}_\perp\,' \\
                       \dd \vec{q}_\perp \end{pmatrix} \, .
\end{equation}
The relations between the submatrices of $N$ and those of $M$
are given by:
$\tilde{A} = - B^{-1} A$, $\tilde{B} = B^{-1}$,
$\tilde{C} = C - D \, B^{-1} A$, and $\tilde{D} = D \, B^{-1}$.
On the other hand, the matrix elements of $N$ can be obtained
from the action function of the trajectory $S(\vec{q},\vec{q}\,',E)$
through
\begin{equation}
N_{ij} = \sigma_i \,\frac{\partial^2 S}{\partial z_i \partial z_j}
\, , \qquad \qquad \vec{z} = \begin{pmatrix}
\dd \vec{q}_\perp\,' \\ \dd \vec{q}_\perp \end{pmatrix}
\, , \qquad \qquad
\sigma_i = \begin{cases} -1 & \text{if $j=1,2$} \\
1 & \text{if $j=3,4$} \end{cases} \, .
\end{equation}
Since the mixed second derivatives of the action function do
not depend on the order of the derivation
this yields 6 conditions for the matrix elements
of $N$ which are equivalent to $\tilde{A} = \tilde{A}^T$,
$\tilde{B} = -\tilde{C}^T$, and $\tilde{D} = \tilde{D}^T$.

\section{Determination of the stability matrix}
\label{apb}

In two-dimensional billiard systems the stability matrix
is composed of two kinds of partial stability matrices
$M_t$ and $M_r$ where $M_t$ denotes the matrix for the free
motion between two reflections and $M_r$ is the matrix
for a reflection. They have the following form (see \cite{SS90b})
\begin{equation} \label{apb1}
M_t = \begin{pmatrix} 1 & L/p \\ 0 & 1 \end{pmatrix} \, ,
\qquad \qquad M_r = \begin{pmatrix} -1 & 0 \\
\frac{2 p}{R \cos \alpha} & -1 \end{pmatrix} \, ,
\end{equation}
where $L$ is the length of a trajectory between two
reflections, $R$ is the radius of curvature at a reflection
point, $\alpha$ is the angle of incidence, and $p$
is the modulus of the momentum. 

In three dimensions there is an additional kind of matrix
which describes a rotation of the local coordinate
system around the trajectory \cite{Pri97}. 
The stability matrix for
a trajectory from a point $\vec{r}_a$ to a point $\vec{r}_b$
with $n$ reflections on the boundary can be written in the
following form 
\begin{equation} \label{apb2}
M = M_T^{b \leftarrow n}
M_R^n \, M_S^n
\dots \, M_T^{3 \leftarrow 2} \,
M_R^2 \, M_S^2 \, M_T^{2 \leftarrow 1} \,
M_R^1 \, M_S^1 \, M_T^{1 \leftarrow a} \, .
\end{equation}
Here $M_T$ is the matrix for a part of the trajectory of length $L$
between two reflections. It is a straightforward generalization
of the matrix $M_t$ in two dimensions. $M_S$
corresponds to a rotation of the local coordinate system around the
trajectory such that the new coordinate with
index 2 lies in the reflection plane that is spanned by the
incoming and outgoing trajectory at a reflection point.
The angle of rotation $\theta$ is defined by
\begin{alignat}{4} \label{apb3}
\hat{e}_2 & =   && \hat{e}_2' \cos \theta + \hat{e}_3' \sin \theta \, ,&
\qquad \qquad
\dd q_2 & =   && \dd q_2' \cos \theta + \dd q_3' \sin \theta \, ,
\notag \\
\hat{e}_3 & = - && \hat{e}_2' \sin \theta + \hat{e}_3' \cos \theta \, , &
\qquad \qquad
\dd q_3 & = - && \dd q_2' \sin \theta + \dd q_3' \cos \theta \, ,
\end{alignat}
where $\hat{e}_i'$ are the unit vectors along the old coordinate
axes and $\hat{e}_i$ are the unit vectors along the new coordinate
axes. $M_T$ and $M_S$ are given by
\begin{equation} \label{apb4}
M_T = \begin{pmatrix}
1 & 0 & L/p & 0   \\
0 & 1 & 0   & L/p \\
0 & 0 & 1   & 0   \\
0 & 0 & 0   & 1
\end{pmatrix}
\; , \qquad
M_S = \begin{pmatrix}
 \cos \theta & \sin \theta & 0 & 0 \\
-\sin \theta & \cos \theta & 0 & 0 \\
0 & 0 &  \cos \theta & \sin \theta \\
0 & 0 & -\sin \theta & \cos \theta 
\end{pmatrix} \; .
\end{equation}
Furthermore, $M_R$ is the matrix for a reflection
\begin{equation} \label{apb5}
M_R = \begin{pmatrix}
-1 & 0 & 0 & 0 \\[10pt] 0 & 1 & 0 & 0 \\[10pt]
 \dfrac{2 p}{R_a \cos \alpha} &
 \dfrac{2 p}{R_c} & -1 & 0 \\[10pt]
-\dfrac{2 p}{R_c} &
-\dfrac{2 p \cos \alpha}{R_b} & 0 & 1
\end{pmatrix} \; .
\end{equation}
Here $\alpha$ is the angle of incidence, i.\,e. the angle
between the incoming (or outgoing) trajectory and the
normal to the boundary. $R_a$ and
$R_b$ are the radii of curvature in the reflection
plane and perpendicular to it, respectively. $R_a$, $R_b$
and $R_c$ can be expressed in terms of the
two main radii of curvature at the reflection point $R_1$
and $R_2$, and the angle $\beta$ between the tangent lying
in the reflection plane and the direction of the main
curvature $1/R_1$ (see eq.\ (\ref{apd2})).
\begin{equation}  \label{apd4}
\frac{1}{R_a} = \frac{\cos^2 \beta}{R_1} + \frac{\sin^2 \beta}{R_2}
\; , \qquad
\frac{1}{R_b} = \frac{\sin^2 \beta}{R_1} + \frac{\cos^2 \beta}{R_2}
\; , \qquad
\frac{1}{R_c} = \frac{R_2 - R_1}{R_1 R_2 } \cos \beta \, \sin \beta \; .
\end{equation}

The form of the matrices $M_T$, $M_S$ and $M_R$ has been
given in $\cite{Pri97}$ for the case of reflections on
spheres ($R_1 = R_2$).
In $\cite{Wir95}$ these results were generalized to
cases where one of the main curvature directions of the billiard
boundary lies in the reflection plane ($\theta = 0$).
Here we give the form of $M_R$ for arbitrary reflections.
The derivation is carried out in appendix \ref{apd}.

Finally, we discuss the stability matrix for a periodic orbit.
It is given by (\ref{apb2}) if $a$ is replaced by $n$ and
the term $M_T^{b \leftarrow n}$ is dropped (see eq.\ (\ref{apc1})). 
In this form it is assumed that the local coordinate
system after one traversal of the trajectory has the same
orientation as it has at the beginning. Otherwise an additional
rotation matrix $M_S$ has to be multiplied to the product.
However, since one has the freedom to choose
the orientation of the local coordinate system at the starting
point, one can always choose it to have the same orientation
as at the end point. In other words, for the evaluation
of (\ref{apc1}) from right to left one starts with a local
coordinate systems whose second coordinate lies in the
reflection plane at $\vec{r}_n$.

\section{Determination of the Maslov index}
\label{apc}

We summarize in this appendix the results of sections
\ref{secc} and \ref{secd} for the Maslov index of a
trajectory. Consider a trajectory with $n$ reflections
on the billiard wall. If one of the reflection points
is chosen as starting point of the trajectory the
stability matrix can be written in the following form
\begin{equation} \label{apc1}
M = M_R^n \, M_S^n \, M_T^{n \leftarrow n-1}
\dots \, M_T^{3 \leftarrow 2} \,
M_R^2 \, M_S^2 \, M_T^{2 \leftarrow 1} \,
M_R^1 \, M_S^1 \, M_T^{1 \leftarrow n} \, .
\end{equation}
Here the orientation of the local coordinate system 
at the initial point has to agree with its orientation
at the final point as discussed in the appendix~\ref{apb}.
Now the trajectory is traversed once, and correspondingly the
multiplication of partial matrices in (\ref{apc1}) 
is carried out. The Maslov index changes by two for
every reflection matrix in this product (in case of Dirichlet
boundary conditions). Otherwise it changes only during a
part of the trajectory between two reflections. How
much it changes is determined by the following method.
For every $M_T$-matrix in (\ref{apc1}) except for 
the first one $M_T^{1 \leftarrow n}$ one
denotes the product of partial matrices
up to this $M_T$-matrix by $M'$, and the product including
this $M_T$-matrix by $M$ such that $M = M_T \, M'$.
Then one considers the following two quantities 
\begin{equation}
Z_1 = \frac{M_{13}  \, M_{24}' - M_{14}  \, M_{23}'}{
            M_{13}' \, M_{24}' - M_{14}' \, M_{23}'}
\, , \qquad \qquad
Z_2 = \frac{M_{13}  \, M_{24}  - M_{14}  \, M_{23} }{
            M_{13}' \, M_{24}' - M_{14}' \, M_{23}'} \, .
\end{equation}
The Maslov index changes according to the following rule
\begin{equation}
\nu = \nu' +
\begin{cases}
0 & \text{if $Z_1 > 0$ and $Z_2 > 0$} \, , \\
1 & \text{if $Z_2 < 0$}               \, , \\
2 & \text{if $Z_1 < 0$ and $Z_2 > 0$} \, ,
\end{cases}
\end{equation}
where $\nu'$ denotes the previous value of the index.
This follows from (\ref{secc17}) and (\ref{secc20}) by
noting that the $M_{1i}$-element change only their
sign upon a reflection and the $M_{2i}$-elements remain
unchanged. (Note that $M'$ is defined here including
the reflection matrix in contrast to the main section.)

Finally there can be an additional change of the Maslov index
that arises from the evaluation of the trace. It is determined
by considering 
\begin{equation}
Z_1 = \frac{{\cal M}_{42}}{\det B}
\, , \qquad \qquad
Z_2 = \frac{\det(M-1)}{\det B} \, ,
\end{equation}
where $M$ is now the total stability matrix of the periodic orbit
given in (\ref{apc1}), and $B$ is its upper right submatrix as
defined in appendix \ref{apa}. ${\cal M}_{ij}$ is the determinant
of the $3 \times 3$-matrix that is obtained by dropping the
$i$-th row and the $j$-th column of the matrix $(M-1)$.

Then the index $\nu$ is changed according to
\begin{equation}
\mu = \nu +
\begin{cases}
0 & \text{if $Z_1 > 0$ and $Z_2 > 0$} \, , \\
1 & \text{if $Z_2 < 0$}               \, , \\
2 & \text{if $Z_1 < 0$ and $Z_2 > 0$} \, .
\end{cases}
\end{equation}

This yields the total Maslov index $\mu$ of the periodic orbit.

\section{Derivation of the stability matrix for a reflection} 
\label{apd}

Let the point of reflection be at the origin of the coordinate
system. The $x_1'$- and $x_2'$-axes of the coordinate system
are chosen to lie in the directions of the two main curvatures
of the boundary surface. Then the boundary is locally represented by
\begin{equation} \label{apd1}
x_3' = - \frac{{x_1'}^2}{2 R_1} - \frac{{x_2'}^2}{2 R_2} \, ,
\end{equation}
where $R_1$ and $R_2$ are the two main radii of curvature.
Now the coordinate system is rotated about the $x_3'$-axis
so that the new $x_1$-axis lies in the plane of reflection,
and the component of the incoming momentum in $x_1$-direction
is non-negative
\begin{equation} \label{apd2}
x_1' =   x_1 \cos \beta + x_2 \sin \beta \, , \qquad 
x_2' = - x_1 \sin \beta + x_2 \cos \beta \, , \qquad
x_3' =   x_3 \, .
\end{equation}
In the new coordinates the boundary is locally represented by
\begin{equation} \label{apd3}
x_3 = - \frac{x_1^2}{2 R_a} - \frac{x_2^2}{2 R_b}
      - \frac{x_1 x_2}{R_c} \, ,
\end{equation}
where $R_a$ and $R_b$ are the radii of curvature in direction of
the new coordinate axes in planes containing the normal vector,
and the relations between $R_a$, $R_b$, $R_c$ and $R_1$, $R_2$
and $\beta$ are given in (\ref{apd4}).
\bigskip

Quantities before the reflection are given a prime in the following,
and the conserved absolute value of the momentum is denoted by $p$.
The angle of incidence $\alpha$ is defined
as the angle between the incoming momentum $\vec{p}\,'$ and the
outward pointing normal vector to the boundary $\hat{n}$,
$0 \leq \alpha < \pi/2$.
The outgoing momentum is $\vec{p}$. One has
\begin{equation} \label{apd5}
\hat{n} = \left( \begin{array}{c} 0 \\ 0 \\ 1 \end{array} \right)
\; , \quad
\vec{p}\,' = p \left( \begin{array}{c} 
\sin \alpha \\ 0 \\ \cos \alpha \end{array} \right)
\; , \quad
\vec{p} = \vec{p}\,' - 2 (\vec{p}\,' \cdot \hat{n}) \hat{n}
=  p \left( \begin{array}{c} 
\sin \alpha \\ 0 \\ - \cos \alpha \end{array} \right) \; .
\end{equation}
In the following one considers an infinitesimal change in the
position and the momentum of the incoming trajectory and
determines the corresponding deviations for the outgoing
trajectory. First we discuss how the normal
vector changes if the point of reflection is changed.
The new normal vector is determined from two tangential
vectors at the point $\vec{r} = (x_1, x_2, x_3)$ with $x_3$
given by eq.\,(\ref{apd3})
\begin{equation} \label{apd6}
\vec{t}_1 = \frac{\partial \vec{r}}{\partial x_1} = \begin{pmatrix} 
1 \\[6pt] 0 \\[6pt] -\dfrac{x_1}{R_a} -
\dfrac{x_2}{R_c} \end{pmatrix} \, , \qquad
\vec{t}_2 = \frac{\partial \vec{r}}{\partial x_2} = \begin{pmatrix}
0 \\[6pt] 1 \\[6pt] -\dfrac{x_2}{R_b} -
\dfrac{x_1}{R_c} \end{pmatrix} \; ,
\end{equation}
from which one obtains
\begin{equation} \label{apd7}
\hat{n} = \dfrac{\vec{t}_1 \times \vec{t}_2}{|\vec{t}_1 \times \vec{t}_2|}
= \begin{pmatrix} \dfrac{x_1}{R_a} +
\dfrac{x_2}{R_c} \\[10pt] \dfrac{x_2}{R_b} +
\dfrac{x_1}{R_c} \\[10pt] 1 \end{pmatrix} \,
\left[ 1 + \left( \dfrac{x_1}{R_a} + \dfrac{x_2}{R_c} \right)^2
         + \left( \dfrac{x_2}{R_b} + \dfrac{x_1}{R_c} \right)^2
\right]^{-1/2} \; .
\end{equation}
Now one considers the four cases that one of the infinitesimal
quantities $\dd q_2'$, $\dd q_3'$, $\dd p_2'$ and $\dd p_3'$ is different
from zero. They describe infinitesimal deviations from
the incoming trajectory in position and momentum, respectively,
immediately before the reflection. 
They are expressed in a local coordinate system, where the first
coordinate is parallel to the trajectory and the other two coordinates
are perpendicular to it.
The quantities with index 2 are in the plane of reflection and
those with index 3 perpendicular to it. 
For each of these four cases the corresponding deviations
for the outgoing trajectory immediately after the reflection are
determined. They are written without a prime.

\begin{description}
\item{\underline{$\dd p_2' \neq 0$:}} 
One sees easily that only $\dd p_2$ is different from zero: $\dd q_2=0$,
$\dd q_3=0$, $\dd p_2=-\dd p_2'$ and $\dd p_3=0$.
\item{\underline{$\dd p_3' \neq 0$:}}
Now only $\dd p_3$ is different from zero: $\dd q_2=0$, $\dd q_3=0$,
$\dd p_2=0$ and $\dd p_3=\dd p_3'$.
\item{\underline{$\dd q_2' \neq 0$:}}
The deviation of the point of reflection is given by
$\dd x_1 = \dd q_2' / \cos \alpha = - \dd q_2 / \cos \alpha$ and
$\dd x_2 = 0$. It follows that $\dd q_2 = - \dd q_2'$ and $\dd q_3=0$.
With the normal vector in (\ref{apd7}), and with $\vec{p}\,'$ and the
relation for $\vec{p}$ in (\ref{apd5}) one finds:
\begin{equation} \label{apd8}
\vec{n} = \begin{pmatrix}
\dfrac{\dd q_2'}{R_a \cos \alpha} \\[10pt]
\dfrac{\dd q_2'}{R_c \cos \alpha} \\[10pt] 1 \end{pmatrix}
\; , \quad
\vec{p} = p \begin{pmatrix}
  \sin \alpha - \dfrac{2 \, \dd q_2'}{R_a} \\[10pt]
- \dfrac{2 \, \dd q_2'}{R_c} \\[10pt]
- \cos \alpha - \dfrac{2 \tan \alpha \, \dd q_2'}{R_a} 
\end{pmatrix} = \begin{pmatrix}
p \sin \alpha - \cos \alpha \dd p_2 \\[6pt] \dd p_3 \\[6pt]
- p \cos \alpha - \sin \alpha \dd p_2
\end{pmatrix} \; .
\end{equation}
From this follows that $\dd p_2 = 2 \, p \, \dd q_2' / (R_a \cos \alpha)$
and  $\dd p_3 = - 2 \, p \, \dd q_2' / R_c$.
\item{\underline{$\dd q_3' \neq 0$:}}
The deviation of the point of reflection
is given by $\dd x_1 = 0$ and $\dd x_2 = \dd q_3' = \dd q_3$, and it
follows that $\dd q_2 = 0$ and $\dd q_3 = \dd q_3'$. With the normal
vector in (\ref{apd7}), and with $\vec{p}\,'$ and the relation for
$\vec{p}$ in (\ref{apd5}) one finds:
\begin{equation} \label{apd9}
\vec{n} = \begin{pmatrix} \dfrac{\dd q_3'}{R_c} \\[10pt]
\dfrac{\dd q_3'}{R_b} \\[10pt] 1 \end{pmatrix}
\; , \qquad
\vec{p} = p \begin{pmatrix}
  \sin \alpha - \dfrac{2 \cos \alpha \, \dd q_3'}{R_c} \\[10pt]
- \dfrac{2 \cos \alpha \, \dd q_3'}{R_b} \\[10pt]
- \cos \alpha - \dfrac{2 \sin \alpha \, \dd q_3'}{R_c} 
\end{pmatrix}
= \begin{pmatrix}
  p \sin \alpha - \cos \alpha \, \dd p_2 \\[6pt] \dd p_3 \\[6pt]
- p \cos \alpha - \sin \alpha \, \dd p_2 \end{pmatrix} \, .
\end{equation}
From this follows that $\dd p_2 = 2 \, p \, \dd q_3' / R_c$
and  $\dd p_3 = - 2 \, p \cos \alpha \, \dd q_3' / R_b$.
\end{description}
Collecting all results one obtains the form of the stability
matrix for a reflection in (\ref{apb5}).

\bibliographystyle{my_unsrt}
\bibliography{../Bifurc/Tex/paper}

\end{document}